\documentstyle[aps,preprint]{revtex}
\tighten
\begin{document}
\preprint{
\parbox{4cm}{
\baselineskip=12pt
TMUP-HEL-9901\\
\vspace*{1cm}}}
\title{Determination of the dynamically generated Yukawa coupling\\
       in supersymmetric QCD}
\author{Noriaki Kitazawa\thanks{e-mail: kitazawa@phys.metro-u.ac.jp}}
\address{Department of Physics, Tokyo Metropolitan University,\\
         Hachioji, Tokyo 192-0397, Japan}
\date{\today}
\maketitle
\begin{abstract}
The strength of the dynamically generated Yukawa coupling
 among composite fields is calculated.
The system of
 $N=1$ supersymmetric SU(2) gauge theory with massive three flavors
 is considered as an example.
We use the techniques of
 ``integrating in'' the gluino-gluino bound state
 in the low energy effective theory
 and the instanton calculation and Shifman-Vainshtein-Zakharov sum rule
 (QCD sum rule) in the fundamental theory.
The obtained value of the Yukawa coupling is of the order of unity.
The method which is developed in this paper
 can be applied to the other supersymmetric gauge theories.
\end{abstract}
\newpage

\section{Introduction}
\label{sec:intro}

Recent development of the techniques
 for analyzing supersymmetric gauge theories \cite{exact-W}
 arises the revival of the investigation
 of supersymmetric composite models
 \cite{NS,LM,KLS,KO,H,HO,LT}.
One of the reason of the revival is that
 the techniques allow us to get
 not only the particle contents at low energy,
 but also the dynamically generated interactions among
 composite particles.
In many models
 the dynamically generated Yukawa interactions
 are identified with or related to the Yukawa interactions
 among Higgs and quarks or leptons in the standard model.
However, the strength of the interactions
 is not satisfactorily determined yet.
In many cases
 one assumes that it is of the order of unity,
 but, on the other hand, there is a claim that
 it must be of the order of $4\pi$ \cite{NDA}.
Some explicit calculations on the dynamics
 are required to determine the strength,
 since it includes the information of the K\"ahler potential
 which can not be determined only by the symmetry and holomorphy.

Naive dimensional analysis (NDA) of Ref.\cite{NDA}
 is the first attempt to determine the coupling constants
 in the low energy effective theories
 of supersymmetric gauge theories.
The strength of coupling constants,
 especially for Yukawa couplings,
 are determined by the renormalization
 from the Seiberg's effective fields
 to the canonically normalized effective fields.
In NDA the renormalization factor is determined
 by assuming that the magnitude of the one-loop correction
 in the effective theory is comparable
 with the tree-level contribution,
 and get the Yukawa coupling of the order of $4\pi$.
This criterion is effective in the chiral Lagrangian for real QCD.
In fact the NDA value of the pion-nucleon Yukawa coupling, $4\pi$,
 is close to the experimental value, $13.5$ \cite{ericson}. 

In this paper we determine the strength
 of the dynamically generated Yukawa coupling among composite fields
 by doing an explicit calculation in the fundamental gauge theory.
We consider
 $N=1$ supersymmetric SU($N_c=2$) gauge theory
 with $N_f=3$ massive flavors as an example.
In the next section
 the relation between the dynamically generated Yukawa coupling
 and the normalization of the effective field is discussed.
The argument is almost the same
 with which has been given in Ref.\cite{NDA}.
We calculate the squark pair condensate
 as a function of $\Lambda$ and $g_Y$,
 the scale of dynamics in the effective theory
 and the Yukawa coupling, respectively,
 and compare it with the result given by the instanton calculation
 in the fundamental theory.
Since the result of the instanton calculation is described by
 the scale of dynamics in the fundamental theory
 $\Lambda_{N_c,N_f} = \Lambda_{2,3}$,
 the Yukawa coupling, $g_Y$, is described by the ratio of
 $\Lambda/\Lambda_{2,3}$.
In Section \ref{sec:glu-eff}
 the chiral superfield of the gluino-gluino bound state
 is introduced in the effective theory
 using the technique of ``integrating in'' \cite{int-in},
 and the mass of the bound state is calculated.
In Section \ref{sec:glu-fun}
 a condition which the mass of the bound state follows
 is obtained using Shifman-Vainshtein-Zakharov (SVZ) sum rule
 (QCD sum rule) \cite{SVZ} in the fundamental theory.
Then,
 we estimate the ratio of $\Lambda/\Lambda_{2,3}$
 using the result of the previous section,
 and obtain a numerical value of the Yukawa coupling.
The resultant value is $g_Y \simeq 0.5 \sim 1$.
In the last section we give a summary and conclude.

\section{Dynamically generated Yukawa coupling}
\label{sec:yukawa}

The Lagrangian of the fundamental theory,
 $N=1$ supersymmetric SU(2) gauge theory with massive three flavors,
 is written as follows.
\begin{eqnarray}
 {\cal L} &=& - \int d^4 \theta \ Q^{\dag i} e^{-2 g_0 {\cal V}} Q_i
              + \int d^2 \theta \ {1 \over 2} m_0 \
                     J^{ij} \epsilon_{\alpha\beta}
                     Q^\alpha_i Q^\beta_j + \mbox{h.c.}
 \nonumber\\
          & & + {1 \over 4} \int d^2 \theta \
                     W^{a \dot{\alpha}} W^a_{\dot{\alpha}}
                     + \mbox{h.c.} 
\end{eqnarray}
Here,
 $Q^\alpha_i$ is the quark chiral superfield,
 ${\cal V}$ is the gluon vector superfield,
 $W^{a \dot{\alpha}}$ is the gluon field strength chiral superfield,
 $g_0$ is the bare gauge coupling constant,
 and $m_0$ is the bare quark mass (flavor independent).
The indices $\alpha, \beta = 1,2$ and $a = 1,2,3$
 are of the fundamental and adjoint representations
 for SU(2) gauge group, respectively,
 $i,j = 1,2, \cdots, 6$ are the flavor indices,
 and $J = \mbox{diag}(\epsilon, \epsilon, \epsilon)$
 is the Sp(3) invariant matrix.
See Appendix for notations.
The confinement is expected at low energy,
 and the effective field
\begin{equation}
 V_{ij} \sim \epsilon_{\alpha\beta} Q^\alpha_i Q^\beta_j
\end{equation}
 is expected to describe the lightest bound state
 by 't Hooft anomaly matching conditions \cite{matching},
 where $V$ is the canonically normalized field with dimension one.
Moreover,
 it is well known that the effective field follows the superpotential
\begin{equation}
 \tilde{W}_{eff}
  = - {1 \over {\Lambda_S^3}} \mbox{Pf} \ \tilde{V}
    - {1 \over 2} m \ \mbox{tr} \left(J \tilde{V} \right) 
\end{equation}
 in the lowest order in the derivative expansion \cite{exact-W}.
Here $\tilde{V}$, which is proportional to the effective field $V$,
 is the Seiberg's effective field with dimension two
 and directly related to the operator
 $\epsilon_{\alpha\beta} Q^\alpha_i Q^\beta_j$
 in the fundamental theory.
The renormalization-group invariant quark mass parameter $m$
 in the low energy effective theory
 is proportional to the renormalized quark mass
 in the fundamental theory.
The first term of the above superpotential is the Yukawa interaction
 \footnote{
 If $m$ is kept finite,
  it describes the Yukawa interactions among massive composite fields.
 To have the Yukawa interaction among massless composite fields,
  we have to set $m$ to zero and introduce some gauge interactions
  by which the origin of the moduli space is chosen\cite{NS,KO}}.

Although the K\"ahler potential can not be determined exactly,
 we can expect
\begin{equation}
 \tilde{K}_{eff}
 = {a \over {\Lambda_S^2}}
   {1 \over 2} \mbox{tr} \left( \tilde{V}^{\dag} \tilde{V} \right)
\end{equation}
 with a positive coefficient $a$
 in the lowest order in the derivative expansion
 by assuming that the effective field $\tilde{V}$ propagates
 without its vacuum expectation value.
The effective action is obtained
 from the following effective Lagrangian.
\begin{equation}
 {\cal L}_{eff} = - \int d^4 \theta \ \tilde{K}_{eff}
                  + \left(
                     \int d^2 \theta \ \tilde{W}_{eff}
                     + \mbox{h.c.}
                    \right).
\label{L-seiberg}
\end{equation}

Since the theory has unique scale of the dynamics,
 all the couplings and coefficients in the effective Lagrangian
 should become of the order of unity,
 if all dimensionful quantities are scaled appropriately \cite{NDA}.
In fact, if we scale
\begin{equation}
 \hat{V} = \left( {\Lambda \over F} \right)^2 \tilde{V},
\quad
 \hat{\theta} = \theta \Lambda^{1/2},
\quad
 \hat{\bar{\theta}} = \bar{\theta} \Lambda^{1/2}
\quad
 \mbox{and}
\quad
 \hat{m} = {m \over \Lambda},
\label{renorm}
\end{equation}
 then the effective Lagrangian becomes
\begin{equation}
 {\cal L}_{eff} = F^2
  \left\{ - \int d^4 \hat{\theta} \ \hat{K}_{eff}
          + \left(
             \int d^2 \hat{\theta} \ \hat{W}_{eff}
             + \mbox{h.c.}
            \right)
  \right\}
\label{L-nondim}
\end{equation}
 with
\begin{eqnarray}
 \hat{K}_{eff}
  &=& {1 \over 2} \mbox{tr} \left( \hat{V}^{\dag} \hat{V} \right), \\
 \hat{W}_{eff}
  &=& - \mbox{Pf} \ \hat{V}
    - {1 \over 2} \hat{m} \ \mbox{tr} \left(J \hat{V} \right).
\end{eqnarray}
Here, $\Lambda = \Lambda_S / a^2$ and $F = \Lambda_S / a^{5/2}$.

We can determine the canonically normalized effective field
 by imposing that the coefficient of the kinetic term is unity.
Namely,
\begin{equation}
 V = {F \over \Lambda} \hat{V} = {\Lambda \over F} \tilde{V},
\end{equation}
 and
\begin{equation}
 {\cal L}_{eff} =
          - \int d^4 \theta \ K_{eff}
          + \left(
             \int d^2 \theta \ W_{eff} + \mbox{h.c.}
            \right)
\label{L-canoni}
\end{equation}
 with
\begin{eqnarray}
 K_{eff} &=& {1 \over 2} \mbox{tr} \left( V^{\dag} V \right), \\
 W_{eff} &=& - g_Y \mbox{Pf} \ V
             - {1 \over 2} {\Lambda \over {g_Y}}
               m \ \mbox{tr} \left(J V \right),
\end{eqnarray}
 where $g_Y \equiv \Lambda^2/F = a^{-3/2}$
 is nothing but the Yukawa coupling.

Note that the scale $\Lambda$ in Eq.(\ref{renorm})
 does not necessary coincide with $\Lambda_S$.
If we may set $\Lambda = \Lambda_S$,
 we have $a=1$ and $g_Y = 1$.
This is the result of too strong requirement
 that all couplings and coefficients should become of the order of unity
 by the scaling of Eq.(\ref{renorm})
 with $\Lambda_S$ instead of $\Lambda$.. 

In NDA
 the Yukawa coupling $g_Y$ is determined
 under the requirement that the one-loop quantum effect
 in the Lagrangian of Eq.(\ref{L-nondim})
 is the same order of the tree-level effect.
Namely, when $\hat{m} < 1$ (light matter), the requirement is
\begin{equation}
 {{\Lambda^4} \over {(4\pi)^2 F^2}} \simeq 1,
\end{equation}
 where $(4\pi)^2 F^2$ is the one-loop suppression factor
 and $\Lambda$ is introduced as the ultraviolet cutoff
 \footnote{Note that $\Lambda=1$
 in the Lagrangian of Eq.(\ref{L-nondim}),
 since the unit of the energy is $\Lambda$.}.
Then, we have $g_Y \simeq 4\pi$ for small $m < \Lambda$.

The squark pair condensate is obtained
 using the effective Lagrangian of Eq.(\ref{L-canoni}).
From the supersymmetric vacuum condition
\begin{equation}
 {{\partial W_{eff}} \over {\partial V_{ij}}} = 0
\end{equation}
 and the assumption of $\langle V_{ij} \rangle = v J_{ij}$,
 we obtain
\begin{equation}
 v = \pm {\sqrt{m \Lambda} \over {g_Y}}.
\end{equation}
Therefore, we have
\begin{equation}
 \langle
  m_0 \epsilon_{\alpha\beta} A_Q^\alpha{}_{i=1} A_Q^\beta{}_{j=2}
 \rangle
 = m \langle \tilde{V}_{12} \rangle
 = m \left( {F \over \Lambda} v \right)
 = \pm {\sqrt{m^3 \Lambda^3} \over {g_Y^2}},
\end{equation}
 where $A_Q^\alpha{}_i$ is the squark field.
This is a renormalization-group invariant quantity.
The same result is obtained from the condition of
 $\partial \tilde{W}_{eff} / \partial \tilde{V} = 0$.
The gluino pair condensate
 is also obtained through Konishi anomaly \cite{konishi}.
\begin{equation}
 \langle
  {{g_0^2} \over {32\pi^2}}
  \lambda^{a \dot{\alpha}} \lambda^a_{\dot{\alpha}}
 \rangle
 =
 \langle
  m_0 \epsilon_{\alpha\beta} A_Q^\alpha{}_{i=1} A_Q^\beta{}_{j=2}
 \rangle
 = \pm {\sqrt{m^3 \Lambda^3} \over {g_Y^2}},
\label{cond-eff}
\end{equation}
 where $\lambda^a_{\dot{\alpha}}$ is the gluino field.
This is also a renormalization-group invariant quantity.

The gluino pair condensate has already been reliably estimated
 by the instanton calculation for $N=1$ supersymmetric
 SU($N_c$) gauge theories with $N_f$ flavors \cite{amati-et-al}
 \footnote{
  It is known that this instanton calculation gives incorrect
   numerical coefficients\cite{incorr-coeff}.
  However, it does not affect the result of this paper,
   since the difference is a factor of the order of unity
   in the case of $SU(2)$ gauge group.}.
\begin{equation}
 \langle
  {{g_0^2} \over {32\pi^2}}
  \lambda^{a \dot{\alpha}} \lambda^a_{\dot{\alpha}}
 \rangle
 =
 \left( C_{N_c}
        \left( \Lambda_{N_c,N_f}^{\rm 1-loop} \right)^{3N_c-N_f}
        \left( 1 + {\cal O}(g(\mu)^4) \right)
        {1 \over {g(\mu)^{2N_c}}}
        \prod_{i=1}^{N_f} m_i(\mu)
 \right)^{1/N_c}
 \ \mbox{e}^{2 \pi i k / N_c},
\end{equation}
 where $k = 1,2, \cdots, N_c$,
 the scale $\Lambda_{N_c,N_f}^{\rm 1-loop}$ is the one
 where the one-loop running coupling diverges,
 $g(\mu)$ and $m_i(\mu)$ are the renormalized coupling
 and mass, respectively,
 and $C_{N_c} \equiv 2^{2N_c} / (N_c-1)! (3N_c-1)$.
This result is obtained
 by evaluating the one-loop quantum fluctuation
 around the single instanton background,
 and the reliability of the approximation is guaranteed
 by the supersymmetric Ward-Takahashi identities.
In the above equation
 ${\cal O}(g(\mu)^4)$ indicates the contribution
 from the higher-loop quantum fluctuation.
We can rewrite this quantity as follows \cite{amati-et-al}.
\begin{eqnarray}
 & &
 \left( \Lambda_{N_c,N_f}^{\rm 1-loop} \right)^{3N_c-N_f}
 \left( 1 + {\cal O}(g(\mu)^4) \right)
 {1 \over {g(\mu)^{2N_c}}}
 \prod_{i=1}^{N_f} m_i(\mu)
\nonumber\\
 &=&
 \mu^{3N_c-N_f}
 \exp \left\{
       - {{8\pi^2} \over {g(\mu)^2}}
       \left( 1 + {\cal O}(g(\mu)^2) \right)
      \right\}
 {1 \over {g(\mu)^{2N_c}}}
 \prod_{i=1}^{N_f} m_i(\mu)
\nonumber\\
 &=&
 \mu^{3N_c-N_f}
 \exp \left(
       -(3N_c-N_f) \int_g^{g(\mu)} {{dg'} \over {\beta(g')}}
      \right)
 \exp \left(
       -N_f \int_g^{g(\mu)} dg' {{\gamma_m(g')} \over {\beta(g')}}
      \right)
 \prod_{i=1}^{N_f} m_i(\mu)
\nonumber\\
 &=&
 \left( \Lambda_{N_c,N_f} \right)^{3N_c-N_f}
 \prod_{i=1}^{N_f} [m_i]_{\rm inv},
\end{eqnarray}
 where $\beta(g)$ is the $\beta$-function \cite{exact-beta}
\begin{equation}
 \beta(g) = - {{g^3} \over {16\pi^2}} \cdot
              {{3N_c - N_f + N_f \gamma_m(g)}
               \over
               {1-N_c g^2 / 8\pi^2 + {\cal O}(g^4)}}
\end{equation}
 and $\gamma_m(g)$ is the anomalous dimension of mass.
The renormalization-group invariant quantities
 $\Lambda_{N_c,N_f}$ and $[m_i]_{\rm inv}$ are defined as
\begin{equation}
 \Lambda_{N_c,N_f}
  = \mu \exp \left(
              - \int_g^{g(\mu)} {{dg'} \over {\beta(g')}}
             \right),
\label{non-pert-scale}
\end{equation}
\begin{equation}
 [m_i]_{\rm inv}
  = m_i(\mu)
    \exp \left(
          - \int_g^{g(\mu)} dg' {{\gamma_m(g')} \over {\beta(g')}}
         \right),
\end{equation}
 where $g$ satisfies
\begin{equation}
 g^{2N_c} \exp \left(
                 {{8\pi^2} \over {g^2}} (1 + {\cal O}(g^2))
               \right) = 1.
\end{equation}
Therefore,
 in case of $N_c=2$ and $N_f=3$ and that all masses are degenerate
 we have
\begin{equation}
 \langle
  {{g_0^2} \over {32\pi^2}}
  \lambda^{a \dot{\alpha}} \lambda^a_{\dot{\alpha}}
 \rangle
 = \pm
 \left( C_2
        \left( \Lambda_{2,3} \right)^3
        [m]_{\rm inv}^3
 \right)^{1/2},
\quad
 C_2 = {{16} \over 5}.
\label{cond-fun}
\end{equation}

The mass parameter in the effective theory, $m$,
 can be identified with $[m]_{\rm inv}$,
 since we can consider that the mass term in the effective theory
 is introduced through the replacement of the
 renormalization-group invariant operator
 $m(\mu)
  (\epsilon_{\alpha\beta} Q^\alpha_i Q^\beta_j)_\mu / [m]_{\rm inv}$
 by the effective field $\tilde{V}_{ij}$
 in the superpotential.
Therefore,
 by equating the Eqs. (\ref{cond-eff}) and (\ref{cond-fun})
 we obtain the Yukawa coupling
\begin{equation}
 g_Y =
  \left(
   {1 \over {C_2}} \left( {\Lambda \over {\Lambda_{2,3}}} \right)^3
  \right)^{1/4}
\label{Yukawa}
\end{equation}
 which is the function of the ratio $\Lambda / \Lambda_{2,3}$.
These two scales are not always equal,
 since the scale $\Lambda$ is introduced
 without any concrete relation with the fundamental theory.
The Yukawa coupling can be determined,
 if $\Lambda$ is described by $\Lambda_{2,3}$
 \footnote{If we use the relation
            $\Lambda^3 = \Lambda_S^3 / a^6 = \Lambda_s^3 g_Y^4$,
            Eq.(\ref{Yukawa}) gives just a relation between
            $\Lambda_S$ and $\Lambda_{2,3}$.
           The difference between $\Lambda_S$ and $\Lambda$
            is important.}.
We need another independent quantity which can be calculated
 both in the effective theory and the fundamental theory.
The mass of the gluino-gluino bound state can be the quantity.

\section{Gluino-gluino bound state in the effective theory}
\label{sec:glu-eff}

We introduce the chiral superfield
\begin{equation}
 S \sim
  - {{g_0^2} \over {32\pi^2}} W^{a\dot{\alpha}} W^a_{\dot{\alpha}}
\end{equation}
 whose scalar component is the gluino-gluino bound state
 to the low energy effective theory
 using the method of ``integrating in'' \cite{int-in},
 and calculate its mass.
Following the conjecture of Ref.\cite{int-in},
 we consider the effective superpotential after ``integrating in''
 as follows.
\begin{equation}
 \tilde{W}_{eff}'
  = {\cal G}(\tilde{V},\tilde{S})
    - {1 \over 2} m \ \mbox{tr} \left( J \tilde{V} \right)
    + \ln \Lambda_S^3 \cdot \tilde{S},
\end{equation}
 where $\tilde{S}$ is the Seiberg's effective field
 with dimension three and directly related to the operator
 $- {{g_0^2} \over {32\pi^2}} W^{a\dot{\alpha}} W^a_{\dot{\alpha}}$
 in the fundamental theory.
The conjecture is that
 in the effective superpotential the scale $\Lambda_S$ is included
 only as a coefficient of the field $\tilde{S}$
 with the form of $\ln \Lambda_S^{3N_c-N_f}$.
The function ${\cal G}(\tilde{V},\tilde{S})$ satisfies
\begin{equation}
 {{\partial {\cal G}} \over {\partial \tilde{S}}} = - \ln \Lambda_S^3
\label{legendre-1}
\end{equation}
 due to the supersymmetric vacuum condition
 $\partial \tilde{W}_{eff}' / \partial \tilde{S} = 0$.
On the other hand,
 since $\tilde{W}_{eff}$ is equivalent to $\tilde{W}_{eff}'$
 as the effective superpotential, the relation
\begin{equation}
 {{\partial {\tilde{W}_{eff}}} \over {\partial \ln \Lambda_S^3}}
 =
 {{\partial {\tilde{W}_{eff}'}} \over {\partial \ln \Lambda_S^3}}
 = \tilde{S}
\label{legendre-2}
\end{equation}
 should be satisfied.
This relation gives
\begin{equation}
 \ln \Lambda_S^3 = \ln {{\mbox{Pf} \tilde{V}} \over {\tilde{S}}},
\end{equation}
 and we can integrate Eq.(\ref{legendre-1}) and obtain
\begin{equation}
 {\cal G}(\tilde{V},\tilde{S})
  = \tilde{S}
     \left(
      \ln {{\tilde{S}} \over {\mbox{Pf} \tilde{V}}} - 1
     \right)
  + {\cal F}(\tilde{V}),
\end{equation}
 where ${\cal F}(\tilde{V})$ is a function of $\tilde{V}$.
Therefore, we have
\begin{equation}
 \tilde{W}_{eff}'
  = \tilde{S}
     \left(
      \ln {{\Lambda^3 \tilde{S}} \over {g_Y^4 \mbox{Pf} \tilde{V}}} - 1
     \right)
  - {1 \over 2} m \ \mbox{tr} \left( J \tilde{V} \right)
  + {\cal F}(\tilde{V}),
\label{W-with-S}
\end{equation}
 where the relation
 $\Lambda_S^3 = \Lambda^3 a^6 = \Lambda^3 / g_Y^4$ was used.
This effective superpotential correctly gives
 the gluino pair condensate of Eq.(\ref{cond-eff}).

To obtain the mass of the gluino-gluino bound state,
 the canonically normalized effective field $S$ have to be defined.
We assume the K\"ahler potential
\begin{equation}
 \tilde{K}_{eff}'
 = {a \over {\Lambda_S^2}}
   {1 \over 2} \mbox{tr} \left( \tilde{V}^{\dag} \tilde{V} \right)
 + b \left( \tilde{S}^{\dag} \tilde{S} \right)^{1/3}
\label{assum-K}
\end{equation}
 following Ref.\cite{veneziano-yankielowicz},
 where $b$ is a positive constant.
If the effective field $\tilde{S}$
 is scaled appropriately to the dimensionless one, $\hat{S}$,
 together with the scalings of $\tilde{V}$ to $\hat{V}$ and so on,
 all the couplings and coefficients in the effective Lagrangian
 should become order unity with the overall factor $F^2$.
Since the first term of $\tilde{W}_{eff}'$
 is proportional to $\tilde{S}$, the scaling have to be
\begin{equation}
 \hat{S} = {\Lambda \over {F^2}} \tilde{S}.
\end{equation}
The effective Lagrangian becomes
\begin{equation}
 {\cal L}_{eff} = F^2
  \left\{ - \int d^4 \hat{\theta} \ \hat{K}_{eff}'
          + \left(
             \int d^2 \hat{\theta} \ \hat{W}_{eff}'
             + \mbox{h.c.}
            \right)
  \right\}
\end{equation}
 with
\begin{eqnarray}
 \hat{K}_{eff}'
 &=& {1 \over 2} \mbox{tr} \left( \hat{V}^{\dag} \hat{V} \right)
     + b \left({{\Lambda^2} \over F} \right)^{2/3}
       \left( \hat{S}^{\dag} \hat{S} \right)^{1/3}, \\
 \hat{W}_{eff}'
 &=& \hat{S}
     \left(
      \ln {{\hat{S}} \over {\mbox{Pf} \hat{V}}} - 1
     \right)
  - {1 \over 2} \hat{m} \ \mbox{tr} \left( J \hat{V} \right)
  + {\hat {\cal F}}(\hat{V}).
\end{eqnarray}
The requirement of that
 the coefficient of $( \hat{S}^{\dag} \hat{S} )^{1/3}$
 in $\hat{K}_{eff}'$ is unity gives $b=g_Y^{-2/3}$.

Next,
 we expand $( \tilde{S}^{\dag} \tilde{S} )^{1/3}$
 in $\tilde{K}_{eff}'$ around the vacuum expectation value of
 $\langle \tilde{S} \rangle$
 and define the canonical normalization.
Namely, we set
\begin{equation}
 \tilde{S} = \langle \tilde{S} \rangle + \tilde{S}^q,
\end{equation}
 and get
\begin{equation}
 \tilde{K}_{eff}'
 = {a \over {\Lambda_S^2}}
   {1 \over 2} \mbox{tr} \left( \tilde{V}^{\dag} \tilde{V} \right)
 + {b \over 3} {{\tilde{S}^{q\dag} \tilde{S}^q}
                \over 
                {\left( 
                  \langle \tilde{S}^{\dag} \rangle
                  \langle \tilde{S} \rangle
                 \right)^{2/3}}}
 + \left( 
    \langle \tilde{S}^{\dag} \rangle
    \langle \tilde{S} \rangle
   \right)^{1/3} \cdot
   {\cal O}
   \left(
    \left( 
    \langle \tilde{S}^{\dag} \rangle
    \langle \tilde{S} \rangle
    \right)^{-2}
   \right).
\label{expansion-K}
\end{equation}
Then, the canonically normalized field is defined as
\begin{equation}
 S = \sqrt{{b \over 3}
          {1 \over 
           {\left( 
             \langle \tilde{S}^{\dag} \rangle
             \langle \tilde{S} \rangle
            \right)^{2/3}}}} \ \tilde{S}
   = {{g_Y} \over {\sqrt{3} m \Lambda}} \ \tilde{S}.
\end{equation}
Therefore,
 the mass of the gluino-gluino bound state is obtained as
\begin{equation}
 m_S^2
     = \left|
       \left( {{\sqrt{3} m \Lambda} \over {g_Y}} \right)^2
       \langle
        {{\partial^2 \tilde{W}_{eff}'} \over {\partial \tilde{S}^2}}
       \rangle
       \right|^2
     = \left( {{\sqrt{3} m \Lambda} \over {g_Y}} \right)^4
       {1 \over {|\langle \tilde{S} \rangle|^2}}
     = 9 m \Lambda.
\label{mass-S}
\end{equation}
In the limit of $m \rightarrow \infty$
 the theory becomes supersymmetric SU(2) Yang-Mills theory
 with scale $\Lambda_{\rm SYM} = \sqrt{m \Lambda_S}$,
 and the mass of the gluino-gluino bound state
 is expected to be of the order of $\Lambda_{\rm SYM}$.
Therefore,
 the result of Eq.(\ref{mass-S}) is correct for large $m > \Lambda_S$
 assuming no mass dependence of $g_Y$.
However, it can not be a correct formula for small $m \ll \Lambda_S$,
 since $m_S$ is expected to remain finite
 in the $m \rightarrow 0$ limit with finite $g_Y$.
This means that the assumption of Eq.(\ref{assum-K})
 is not justified for small $m \ll \Lambda_S$.

\section{Gluino-gluino bound state in the fundamental theory}
\label{sec:glu-fun}

We calculate the mass of the gluino-gluino bound state
 using SVZ sum rule (QCD sun rule) \cite{SVZ}
 in the fundamental theory
 \footnote{The mass has already been calculated
           using the similar technique in Ref.\cite{leroy}.}.
The bound state couples to
 both the scalar and auxiliary components of the operator
\begin{equation}
 {\cal O}_S(y,\theta)
  = - {{g_0^2} \over {32\pi^2}}
      W^{a\dot{\alpha}}(y,\theta) W^a_{\dot{\alpha}}(y,\theta)
  = {{g_0^2} \over {32\pi^2}}
      \lambda^{a\dot{\alpha}}(x) \lambda^a_{\dot{\alpha}}(x)
  + \cdots,
\end{equation}
 where $y = x + i \bar{\theta} \sigma \theta$.
Then we consider the quantity
\begin{equation}
 \Pi(Q^2)
  = i \int d^4x e^{iqx}
    \langle
     T \int d^2 \theta \ {\cal O}_S (y,\theta) \ {\cal O}_S (0,0)
    \rangle,
\end{equation}
 where $Q^2 = - q^2$.
This quantity can be described
 in the spectral function representation as
\begin{equation}
 \Pi(Q^2)
  = \int_0^\infty ds \
    {{\rho(s)} \over {s+Q^2-i\epsilon}}
\end{equation}
 with
\begin{equation}
 \rho(s=k^2) \epsilon(k_0)
 = (2\pi)^3 \sum_n \delta^4(p_n-k)
   \langle 0 |
    \int d^2 \theta \ {\cal O}_S (y,\theta) \Big|_{x=0}
   | n \rangle
   \langle n | {\cal O}_S (0,0) | 0 \rangle,
\end{equation}
 where the summation is taken over all the states.
On the other hand,
 $\Pi(q^2)$ can be directly calculated
 in the limit of $Q^2 \rightarrow \infty$
 by the operator product expansion (OPE).
Namely,
\begin{eqnarray}
 &\displaystyle{\lim_{Q^2 \rightarrow \infty}}&
 i \int d^4x e^{iqx}
 T \left\{
    \int d^2 \theta \ {\cal O}_S (y,\theta), {\cal O}_S (0,0)
   \right\}
\nonumber\\
 &=& 2 \left( {{g^2} \over {32\pi^2}} \right)^2
     \displaystyle{\lim_{Q^2 \rightarrow \infty}} i \int d^4x e^{iqx}
 \Bigg[
   T \left\{ {1 \over 4}
              \left(
               v^{a\mu\nu} v^a_{\mu\nu}
               + i v^{a\mu\nu} \tilde{v}^a_{\mu\nu}
              \right)_{(x)}, \
              \left( \lambda^b \lambda^b \right)_{(0)}
     \right\}
\nonumber\\
 && \qquad\qquad\qquad\qquad\quad
 + T \left\{ \left(
              \left(
               \lambda^{\dag}
                i \sigma^\mu \mathop{D}^{\leftarrow}{}_\mu
              \right)^a
              \lambda^a
             \right)_{(x)}, \
             \left( \lambda^b \lambda^b \right)_{(0)}
     \right\}
\nonumber\\
 && \qquad\qquad\qquad\qquad\quad
 + T \left\{ \left( - {{g^2} \over 2}
              \left( A_Q^{\dag} T^a A_Q \right)
              \left( A_Q^{\dag} T^a A_Q \right)
             \right)_{(x)}, \
             \left( \lambda^b \lambda^b \right)_{(0)}
     \right\}
 \Bigg]
\nonumber\\
 &=& A(Q^2) {{g^2} \over {32\pi^2}}
            \left( \lambda^a \lambda^a \right)_{(0)}
\nonumber\\
 &&+ B(Q^2) {1 \over 2} m
            \left(
             J^{ij} \epsilon_{\alpha\beta}
             A^\alpha_Q{}_i A^\beta_Q{}_j
            \right)_{(0)}
\nonumber\\
 &&+ C(Q^2) {{g^2} \over {32\pi^2}}
            \left(
             \lambda^a \lambda^a
             A^{\dag}_Q A_Q
            \right)_{(0)}
\nonumber\\
 &&+ D(Q^2) \left(
            \epsilon_{abc}
            \lambda^a {\bar \sigma}^{\mu\nu} \lambda^b
            v^c_{\mu\nu}
            \right)_{(0)}
\nonumber\\
 &&+ E(Q^2) \left(
            \epsilon_{abc}
            \lambda^a {\bar \sigma}^{\mu\nu} \lambda^b
            {\tilde v}^c_{\mu\nu}
            \right)_{(0)}
\nonumber\\
 &&+ {\cal O}(1/Q^4),
\end{eqnarray}
 where $v^a_{\mu\nu}$ is the gluon field strength
 and $\tilde{v}^a_{\mu\nu}$ is its dual.
All quantities are the renormalized quantities.
Wilsonian coefficients
 $A(Q^2)$, $B(Q^2)$, $C(Q^2)$, $D(Q^2)$ and $E(Q^2)$
 can be determined by the perturbation theory.
Note that the gluino number plus squark number
 (anomalous U(1)${}_R$ symmetry)
 is conserved in the perturbation theory.

By estimating the vacuum expectation values
 of the $T$-products of the both sides 
 multiplied by two $\lambda^{\dag}$'s or two $A_Q^{\dag}$'s
 in the first order of the perturbation theory,
 we obtain
\begin{eqnarray}
 A(Q^2)
  &=& {{\alpha(\mu)} \over {2\pi}}
      \left( 1 + {3 \over {2\pi}} \alpha(\mu)
                 \ln \left( {{Q^2} \over {\mu^2}} \right)
      \right),
\\
 B(Q^2) &=& 0,
\end{eqnarray}
 where $\alpha(\mu) = g(\mu)^2 / 4\pi$.
We consider only the lowest dimensional operators in OPE
 as an approximation.
In the following,
 we take the renormalization point as $\mu = \sqrt{Q^2}$,
 by which the higher order logarithmic correction is suppressed.
Then, we have the sum rule
\begin{equation}
 \int_0^\infty ds {{\rho(s)} \over {s+Q^2-i\epsilon}}
 = - {{\alpha(\sqrt{Q^2})} \over {2\pi}}
   \langle {\cal O}_S(0,0) \rangle
\end{equation}
 for large $Q^2$.
Following Ref.\cite{SVZ},
 we consider the Borel transform of this sum rule.
Namely,
\begin{equation}
 \int_0^\infty ds e^{-s/M^2} \rho(s)
  = - M^2 {{\alpha(\sqrt{M^2})} \over {2\pi}}
      \langle {\cal O}_S(0,0) \rangle,
\label{sum-rule-1}
\end{equation}
 where $M^2$ is a parameter of dimension two
 which corresponds to $Q^2$.
This is the SVZ sum rule in our case.
If there is a value of $M^2$
 which is large enough
 so that $\alpha(\sqrt{M^2})$ in the right hand side is kept small
 and which is small enough so that the integral in the left hand side
 is dominated by the lowest-lying state,
 we can reliably extract the information of the lowest-lying state.
In the following
 we first assume that this is the case,
 and estimate the goodness of the approximation later.

By differentiating the sum rule of Eq.(\ref{sum-rule-1}),
 we obtain
\begin{equation}
 \int_0^\infty ds e^{-s/M^2} s \rho(s)
  = - M^4 {{\alpha(\sqrt{M^2})} \over {2\pi}}
      \langle {\cal O}_S(0,0) \rangle,
\label{sum-rule-2}
\end{equation}
 where we neglect the ${\cal O}(\alpha(\sqrt{M^2})^2)$ term
 in the right hand side.
The ratio of the two sum rules
 of Eqs.(\ref{sum-rule-1}) and (\ref{sum-rule-2}) gives
\begin{equation}
 \int_0^\infty ds e^{-s/M^2} s \rho(s)
 \Bigg/
 \int_0^\infty ds e^{-s/M^2} \rho(s)
 = M^2.
\end{equation}
If the lowest-lying state
 dominates the integrals in the left hand side,
 we can set as
\begin{equation}
 \rho(s=k^2) \simeq
  \delta(k^2-m_S^2)
  \langle 0 |
   \int d^2 \theta \ {\cal O}_S (y,\theta) \Big|_{x=0}
  | k \rangle_S \
  {}_S\langle k | {\cal O}_S (0,0) | 0 \rangle,
\label{one-part-sp}
\end{equation}
 and obtain $M^2 = m_S^2$,
 where $| k \rangle_S$ is the one-particle state of $S$
 with momentum $k$.
Then, the sum rule of Eq.(\ref{sum-rule-1}) becomes
\begin{equation}
 \int_0^\infty ds e^{-s/m_S^2} \rho(s)
  = - m_S^2 {{\alpha(\sqrt{m_S^2})} \over {2\pi}}
      \langle {\cal O}_S(0,0) \rangle.
\label{sum-rule}
\end{equation}

The vacuum expectation value $\langle {\cal O}_S(0,0) \rangle$
 and the matrix elements in the spectral function
 of Eq.(\ref{one-part-sp})
 can be estimated in the effective theory.
It is clear that
\begin{equation}
 \langle {\cal O}_S(0,0) \rangle
 = \langle \tilde{S} \rangle
 = \pm {{\sqrt{m^3 \Lambda^3}} \over {g_Y^2}},
\end{equation}
 and
\begin{equation}
 {}_S\langle k | {\cal O}_S(0,0) | 0 \rangle
 = {}_S\langle k | A_{\tilde{S}}(0) | 0 \rangle
 = {{\sqrt{3} m \Lambda} \over {g_Y}}
   {}_S\langle k | A_S(0) | 0 \rangle
 = {{\sqrt{3} m \Lambda} \over {g_Y}},
\end{equation}
 where $A_{\tilde{S}}$ and $A_S$
 are the scalar components of the effective fields
 $\tilde{S}$ and $S$, respectively.
Moreover,
\begin{equation}
 \langle 0 |
  \int d^2 \theta \ {\cal O}_S (y,\theta) \Big|_{x=0}
 | k \rangle_S
 = \langle 0 | F_{\tilde{S}}(0) | k \rangle_S
 = {{\sqrt{3} m \Lambda} \over {g_Y}}
   \langle 0 | F_S(0) | k \rangle_S,
\end{equation}
 where $F_{\tilde{S}}$ and $F_S$ 
 are the auxiliary components of the effective fields
 $\tilde{S}$ and $S$, respectively.
The auxiliary field $F_S$ can be calculated
 using the effective superpotential of Eq.(\ref{W-with-S}).
\begin{equation}
 F_S = - {{\sqrt{3} m \Lambda} \over {g_Y}}
       \left. {{\partial \tilde{W}_{eff}'{}^{\dag}}
               \over
               {\partial \tilde{S}^{\dag}}} \right|_{\rm scalar}
     = - {{\sqrt{3} m \Lambda} \over {g_Y}}
       \ln {{\sqrt{3} m \Lambda A_S^{\dag}}
            \over
            {g_Y^2 \mbox{Pf} A_V^{\dag}}},
\end{equation}
 where $A_V$ is the scalar component of the effective field $V$.
We expand this expression by $A_S^{\dag}$
 around its vacuum expectation value.
\begin{eqnarray}
 F_S = &-& {{\sqrt{3} m \Lambda} \over {g_Y}}
           {{A_S^{q\dag}} \over {\langle A_S^{\dag} \rangle}}
        +  {\cal O}(1/\langle A_S^{\dag} \rangle^2)
\nonumber\\
       &-& {{\sqrt{3} m \Lambda} \over {g_Y}}
           \ln {{\sqrt{3} m \Lambda \langle A_S^{\dag} \rangle}
                \over
                {g_Y^2 \mbox{Pf} A_V^{\dag}}}.
\end{eqnarray}
The first term describes the coupling with the one-particle state.
Then, we obtain
\begin{equation}
 \langle 0 |
  \int d^2 \theta \ {\cal O}_S (y,\theta) \Big|_{x=0}
 | k \rangle_S
 = - \left( {{\sqrt{3} m \Lambda} \over {g_Y}} \right)^3
     {1 \over {\langle A_{\tilde{S}}^{\dag} \rangle}},
\end{equation}
Therefore, the spectral function can be written as
\begin{equation}
 \rho(s) \simeq
  - \left( {{\sqrt{3} m \Lambda} \over {g_Y}} \right)^4
    {1 \over {\langle \tilde{S} \rangle}}
    \delta(s-m_S^2),
\label{spectral}
\end{equation}
 where we use
 $\langle A_{\tilde{S}}^{\dag} \rangle = \langle \tilde{S} \rangle$.

This result and the sum rule of Eq.(\ref{sum-rule}) give
\begin{equation}
 m_S^2 \alpha(\sqrt{m_S^2})
  = 2\pi \left( {{\sqrt{3} m \Lambda} \over {g_Y}} \right)^4
         {1 \over {\langle \tilde{S} \rangle^2}}
  = 2\pi \cdot 9 m \Lambda.
\end{equation}
Using Eq.(\ref{mass-S}) we have
\begin{equation}
 \alpha(\sqrt{m_S^2}) = 2\pi.
\label{condition}
\end{equation}
This is the condition
 which have to be satisfied
 by the mass of the gluino-gluino bound state.
The expansion parameter on the gauge coupling in the OPE is
\begin{equation}
 {{g(\sqrt{m_S^2})^2} \over {(4\pi)^2}}
 = {{\alpha(\sqrt{m_S^2})} \over {4\pi}}
 = {1 \over 2}.
\end{equation}
This is not much smaller than unity.
However,
 the approximation is enough for the order estimate,
 since the higher order logarithmic correction is suppressed
 by the appropriate selection of the renormalization point.

Now we use the formula of Eq.(\ref{mass-S}).
Since it is reliable only for $m > \Lambda_S$,
 we should not use the running coupling
 for the case of $N_c=2$ and $N_f=3$,
 but the case of $N_c=2$ and $N_f=0$.
Furthermore,
 we have to use the running coupling
 which follows the $\beta$-function \cite{exact-beta}
\begin{equation}
 \beta(\alpha)
 = - {{\alpha^2} \over {2\pi}} \cdot
     {{3N_c}
      \over
      {1 - N_c \alpha/2\pi + {\cal O}(\alpha^2)}},
\qquad
 N_c=2,
\end{equation}
 since the scale of dynamics which is non-perturbativly defined
 by the instanton calculation (see Eq.(\ref{non-pert-scale}))
 has to be introduced.
The solution of the renormalization group equation is
\begin{equation}
 {1 \over {\alpha(\mu)}} + {1 \over \pi} \ln \alpha(\mu)
 = {3 \over \pi} \ln {\mu \over {\Lambda_{2,0}}},
\label{sol-run}
\end{equation}
 where the ${\cal O}(\alpha^2)$ term
 in the denominator of the $\beta$-function
 is neglected as a small contribution.
We can impose the one-loop matching relation,
 $\Lambda_{2,0} = \sqrt{m \Lambda_{2,3}}$
 \footnote{The one-loop matching relation is satisfied
 in the results of the explicit instanton calculation.}.

Now we can determine the value of the ratio $\Lambda/\Lambda_{2,3}$
 using Eqs.(\ref{sol-run}), (\ref{condition}) and (\ref{mass-S}).
\begin{equation}
 {\Lambda \over {\Lambda_{2,3}}}
 = {1 \over 9} (2\pi)^{2/3} e^{1/3} \simeq 0.5.
\label{ratio}
\end{equation}
The scale $\Lambda$ is the same order of $\Lambda_{2,3}$
 as expected.
Now it is possible
 to estimate the magnitude
 of the higher-order operator correction in OPE.
The expansion parameter should be
\begin{equation}
 {{(\Lambda_{2,0})^2} \over {M^2}}
 = {{(\Lambda_{2,0})^2} \over {m_S^2}}
 = {1 \over 9} {{\Lambda_{2,3}} \over \Lambda}
 \simeq 0.2.
\end{equation}
This is small and independent from the mass $m$.
Then, the present approximation is good for the order estimate.

Finally,
 we can determine the value of the Yukawa coupling $g_Y$
 using Eqs.(\ref{ratio}) and (\ref{Yukawa}).
\begin{equation}
 g_Y = \left( 
        {5 \over {16}}
        {{(2\pi)^2 e} \over {9^3}}
       \right)^{1/4}
     \simeq 0.5.
\end{equation}
Namely,
 the resultant value of the dynamically generated Yukawa coupling
 (which is independent from the mass $m$)
 is of the order of unity for large $m > \Lambda_S$,
 which is different from the result of NDA, $4\pi \sim 10$,
 for small $m < \Lambda_S$.

Here, we have to stress that
 the obtained value of the Yukawa coupling is for the theory
 with $m > \Lambda_S$, though it is independent from $m$.
We may consider the simple $m \rightarrow 0$ limit,
 but there are several problems.
For example,
 the mass of the gluino-gluino bound state
 vanishes in this limit (Eq.(\ref{mass-S})),
 which seems to contradict with 't Hooft anomaly matching conditions,
 although the coupling in the spectral function also vanishes
 in this limit (Eq.(\ref{spectral}))
 and the bound state disappears from the spectrum.
To take the massless limit,
 we have to consider the bound state which couples to the operator
 $\mbox{Pf}(\epsilon_{\alpha\beta }A_Q^\alpha A_Q^\beta)$
 in Eq.(\ref{assum-K}), for example.
Since the bound state has the same quantum number of $S$,
 there must be the mixing between them,
 and we can expect that there is no massless bound state
 in the limit of $m \rightarrow 0$, except for $V$.

\section{Conclusion}
\label{sec:conclusion}

The value of the Yukawa coupling
 among the low energy effective fields (composite fields)
 was calculated in the $N=1$ supersymmetric SU(2) gauge theory
 with massive three flavors.
First,
 the value of the squark pair condensate (or gluino pair condensate)
 and the mass of the gluino-gluino bound state
 were calculated in the effective theory
 considering the uniqueness of the scale of dynamics in the theory.
These quantities
 are described by the parameters in the effective theory,
 $\Lambda$, $m$ and $g_Y$.
Next,
 these quantities were evaluated directly in the fundamental theory
 using the technique of the instanton calculation and SVZ sum rule.
The result
 are described by the parameters in the fundamental theory,
 $\Lambda_{2,3}$ and $m$.
Then,
 we obtained the expression of the parameters
 in the effective theory by those of the fundamental theory.
\begin{eqnarray}
 {\Lambda \over {\Lambda_{2,3}}}
 &=& {1 \over 9} (2\pi)^{2/3} e^{1/3} \simeq 0.5, \\
 g_Y
 &=& \left( {5 \over {16}}
            \left(
             {\Lambda \over {\Lambda_{2,3}}}
            \right)^3
     \right)^{1/4}
 \simeq 0.5.
\end{eqnarray}
These results is for large mass $m > \Lambda_S$,
 although they are independent from the mass.
Unfortunately,
 the value can not be directly compared with the result by NDA,
 $g_Y \simeq 4\pi$, for small mass.

We made some approximations in using SVZ sum rule.
The higher order
 in the perturbative gauge coupling in Wilson coefficients
 and the higher-order operator were neglected in the OPE.
The approximations are good for the order estimate,
 since the expansion parameters are not so large:
 $\alpha(\sqrt{m_S^2})/4\pi = 0.5$
 and $\Lambda_{2,0}^2/m_S^2 \simeq 0.2$.
Note that
 the appropriate selection of the renormalization point
 suppresses the higher-order logarithmic correction
 in Wilson coefficients.

The method which is developed in this paper
 can be applied to determine the effective coupling constants
 in the low energy effective theories
 of the other supersymmetric gauge theories.

\acknowledgements

This work was supported in part
 by the Grant-in-Aid for Scientific Research
 from the Ministry of Education, Science and Culture of Japan
 (\#11740156).

\appendix
\section{Notation}
\label{app:notations}

The metric we use is $g = \mbox{diag}(1,-1,-1,-1)$,
 and the $\sigma$-matrices for the two component spinor are
 $(\sigma_\mu)_{\alpha\dot\beta} = (1,\tau^i)$
 and $({\bar\sigma}_\mu)_{\dot\alpha\beta} = (1,-\tau^i)$,
 where $\tau^i$ are the Pauli matrices.
The convention
 on the contraction of the index of two component spinor is
\begin{equation}
 \theta\theta
  = \theta^{\dot\alpha} \theta_{\dot\alpha},
\quad
 {\bar\theta}{\bar\theta}
  = {\bar\theta}^{\alpha} {\bar\theta}_{\alpha},
\end{equation}
 with
 $\theta^{\dot\alpha}
  = \epsilon^{\dot\alpha\dot\beta} \theta_{\dot\beta}$
 and
 ${\bar\theta}^{\alpha}
  = \epsilon^{\alpha\beta} {\bar\theta}_{\beta}$,
 where
 $\epsilon^{\dot\alpha\dot\beta} = \epsilon_{\dot\alpha\dot\beta}$
 and $\epsilon^{\alpha\beta} = \epsilon_{\alpha\beta}$.
The integration over the spinors is defined as
\begin{equation}
 \int d^2 \theta \ \theta^2 = 1,
\quad
 \int d^2 {\bar\theta} \ {\bar \theta}^2 = 1.
\end{equation}
In the followings we give the correspondence
 between the standard notation by Wess and Bagger \cite{wess-bagger}
 and ours.

\newcommand{\th}{\theta}
\newcommand{\thb}{\bar \theta}
\newcommand{\ps}{\psi}
\newcommand{\psb}{\bar \psi}
\newcommand{\ch}{\chi}
\newcommand{\chb}{\bar \chi}
\newcommand{\al}{\alpha}
\newcommand{\ald}{{\dot \alpha}}
\newcommand{\be}{\beta}
\newcommand{\bed}{{\dot \beta}}
\newcommand{\ga}{\gamma}
\newcommand{\gad}{{\dot \gamma}}
\newcommand{\de}{\delta}
\newcommand{\ded}{{\dot \delta}}
\newcommand{\sig}{\sigma}
\newcommand{\sigb}{\bar \sigma}

On the metric and spinors:
\begin{equation}
\eta^{mn} \Big|_{W-B} = - g^{\mu\nu}.
\end{equation}
\begin{equation}
\epsilon^{\al\be} \Big|_{W-B} = \epsilon^{\al\be},
\qquad
\epsilon_{\al\be} \Big|_{W-B} = - \epsilon_{\al\be}.
\end{equation}
\begin{equation}
\left( \sig^m \right)_{\al\bed} \Big|_{W-B}
 = - \left( \sig^\mu \right)_{\al\bed},
\qquad
\left( \sigb^m \right)^{\ald\be} \Big|_{W-B}
 = - \left( \sigb^\mu \right)^{\ald\be}.
\end{equation}
\begin{equation}
\th^\al \Big|_{W-B} = \thb^\al,
\qquad
\thb^\ald \Big|_{W-B} = \th^\ald.
\end{equation}
\begin{equation}
\th\th \Big|_{W-B} = \thb\thb = \thb^\al \thb_\al,
\qquad
\thb\thb \Big|_{W-B} = - \th\th = - \th^\ald \th_\ald.
\end{equation}
\begin{equation}
d^2 \th \Big|_{W-B} = d^2 \thb,
\qquad
d^2 \thb \Big|_{W-B} = - d^2 \th.
\end{equation}

On the chiral superfields:
\begin{equation}
W_\al (y,\th) \Big|_{W-B} = {\bar W}_\al (y^{\dag},\thb),
\qquad
{\bar W}_\ald (y^{\dag},\thb) \Big|_{W-B} = W_\ald (y,\th).
\end{equation}
\begin{equation}
\Phi (y,\th) \Big|_{W-B} = \Phi^{\dag} (y^{\dag},\thb),
\qquad
\Phi^{\dag} (y^{\dag},\thb) \Big|_{W-B} = \Phi (y,\th).
\end{equation}
\begin{equation}
y^m \Big|_{W-B} \equiv x^m + i \th \sig^m \thb \Big|_{W-B}
 = y^{\dag\mu} \equiv x^\mu - i \thb \sig^\mu \th.
\end{equation}

\end{document}